\documentclass[%
,secnumarabic%
,amssymb, amsmath%
, aps%
, pre%
, superscriptaddress%
]{revtex4}
\usepackage{graphicx}

\renewcommand{\d}[1][{\negthickspace}]{\mathrm{d}{#1}\;}

\newcommand{\pdiff}[3][{}]{\frac{\partial^{#1}{#2}}{\partial{#3}^{#1}}}

\renewcommand{\(}[0]{\left (} 
\renewcommand{\)}[0]{\right )}

\newcommand{\llangle}{\left\langle}
\newcommand{\rrangle}{\right\rangle}

\begin{document}

\title{The effect of integration time on fluctuation measurements:\\
  calibrating an optical trap in the presence of motion blur}


\author{Wesley P. Wong}
\email[]{wong@rowland.harvard.edu}
\affiliation{Department of Physics, Harvard University}
\affiliation{Department of Biomedical Engineering, Boston University}
\affiliation{Current: Rowland Institute at Harvard, Harvard University}
\author{Ken Halvorsen}
\altaffiliation{Both authors contributed equally}
\affiliation{Department of Biomedical Engineering, Boston University}
\affiliation{Current: Rowland Institute at Harvard, Harvard University}

\begin{abstract}
Dynamical instrument limitations, such as finite detection bandwidth,
do not simply add statistical errors to fluctuation measurements, but
can create significant systematic biases that affect the measurement
of steady-state properties. Such effects must be considered when
calibrating ultra-sensitive force probes by analyzing the observed
Brownian fluctuations. In this article, we present a novel method for
extracting the true spring constant and diffusion coefficient of a
harmonically confined Brownian particle that extends the standard
equipartition and power spectrum techniques to account for video-image
motion blur. These results are confirmed both numerically with a
Brownian dynamics simulation, and experimentally with laser optical
tweezers.\\
\\ \vskip-.5pc 

\noindent Published in Optics Express, Vol. 14, Issue 25, pp. 12517--12531 (2006).\\
\noindent \small \copyright \, 2006 \hskip.05in
   Optical Society of America \\
\end{abstract}


\maketitle

\section{Introduction}

Investigations of micro- to nano-scale phenomena at finite temperature
(e.g. single-molecule measurements, microrheology) require a detailed
treatment of the Brownian fluctuations that mediate weak interactions
and kinetics \cite{svoboda1994fam,mason1995omf,evans1997dsm,collin05}.
Experimental quantification of such fluctuations are affected by
instrument limitations, which can introduce errors in surprising
ways. Dynamical limitations, such as finite detection bandwidth, do
not simply add statistical errors to fluctuation measurements, but can
create significant systematic biases that affect the measurement of
steady-state properties such as fluctuation amplitudes and probability
densities (e.g. position histograms).

Motion blur, which results from time-averaging a signal over a finite
integration time, can create significant problems when imaging fast
moving objects. It is particularly relevant when measuring the
position fluctuations of a Brownian particle, where even fast
detection methods can have long integration times with respect to the
relevant time scale, as we will demonstrate in this paper. Instrument
bandwidth limitations that arise from motion blur affect a variety of
fluctuation-based measurement techniques, including the quantification
of forces with magnetic tweezers using lateral fluctuations
\cite{strick1996ess}, and microrheology measurements based on the
video-tracking of small particles \cite{chen2003rml}. The issue of
video-image motion blur has recently been addressed in the
single-molecule literature \cite{yasuda1996dmt} and in the field of
microrheology, where the static and dynamic errors resulting from
video-tracking have been carefully analyzed
\cite{savin2005sad,savin2005rfe}. However, discussion has been notably
absent in the area of ultra-sensitive force-probes, despite the
significant effect that it can have on quantitative measurements.
This paper focuses on the practical problem of calibration an optical
trap by analyzing the confined Brownian motion of a trapped particle
\cite{ghislain1993sfm,svoboda1994bao,gittes1998san,florin1998pfm,
bergsoerensen2004psa} in the presence of video-image motion blur.

In this article, we present a novel method for extracting the true
spring constant and diffusion coefficient of a harmonically confined
Brownian particle that extends the standard equipartition and power
spectrum techniques to account for motion blur. In
section\,(\ref{sec:calculation_section}) we describe how the measured
variance of the position of a harmonically trapped Brownian particle
depends on the integration time of the detection apparatus, the
diffusion coefficient of the bead and the trap stiffness. Next, this
theoretical relationship is compared with both simulated data
(section\,(\ref{sec:numerical_studies})) and experimental data using
an optical trap
(section\,(\ref{sec:experimental_verification})). Practical strategies
for trap calibration are given in the discussion
section\,(\ref{sec:discussion}), where we show that motion blur is not
a liability once it is understood, but rather provides valuable
information about the dynamics of bead motion. In particular, we show
how both the spring constant and the diffusion coefficient can be
determined by measuring position fluctuations while varying either the
shutter speed of the acquisition system or the confinement strength of
the trap.

\section{Bias in the measured variance of a harmonically trapped Brownian particle}\label{sec:calculation_section}

Detection systems, such as video cameras and photodiodes, do not
measure the instantaneous position of a particle. Rather, the measured
position $X_m$ is an average of the true position $X$ taken over a
finite time interval, which we call the integration time $W$. In the
simplest model,
\begin{equation}\label{eq:measured_bead_position}
X_m(t) = \frac{1}{W}\int_{t-W}^{t}X(t')\d[t']
\end{equation}
where both the measured and true positions of the particle have been
expressed as functions of time $t$. More complex situations can be
treated by multiplying $X(t')$ by an instrument-dependent function
within the integral, i.e. by using a non-rectangular integration
kernel.

We consider the case of a particle undergoing Brownian motion within a
harmonic potential, $U(x) = \frac{1}{2} k x^2$. In equilibrium, we
expect the probability density of the particle position to be
established by the Boltzmann weight $\exp(-U(x)/k_B T)$, where $k_B$
is the Boltzmann constant and $T$ is the absolute temperature:
\begin{equation}\label{eq:boltzmann_weighting}
\rho_X(x) = \frac{1}{\sqrt{2\pi k_B T/k}} \exp\( - \frac{k x^2}{2 k_B
T}\)
\end{equation}
The variance of the position should then satisfy the equipartition theorem,
\begin{equation} \label{eq:equipartition}
\text{var}(X) \equiv  \llangle X^2 \rrangle - \llangle X \rrangle^2 =
\frac{k_B T}{k}
\end{equation}
However, these equations do not hold for the measured position
$X_m$. In particular, motion blur introduces a systematic bias in the
measured variance,
\begin{equation}
\text{var}(X_m) \le \text{var}(X)
\end{equation}

Following standard techniques (e.g. \cite{oppenheim1996ss}), the
necessary correction can be calculated precisely as a function of the
spring constant $k$, the friction factor of the particle $\gamma$, and
the integration time of the imaging device $W$
\cite{yasuda1996dmt,savin2005sad,savin2005rfe}. First, we define the
dimensionless parameter $\alpha$ by expressing the exposure time $W$
in units of the trap relaxation time $\tau = \gamma / k$, i.e.
\begin{equation}\label{eq:dim_exp_def}
\alpha \equiv \frac{W}{\tau} 
\end{equation}
Note that $\alpha$ can also be expressed in terms of the diffusion
coefficient $D$ by using the Einstein relation $\gamma = k_B T/D$,
i.e. $\alpha = W D k/(k_B T)$. Then as presented in
appendix\,(\ref{appendix:calculation_section}), the measured variance
is given by:
\begin{equation}\label{eq:measuredvar2}
\text{var}(X_m) = \text{var}(X) S(\alpha)
\end{equation}
where $S(\alpha)$ is the motion blur correction function
\begin{equation}\label{eq:universal_blur_fn}
S(\alpha) = \frac{2}{\alpha} - \frac{2}{\alpha^2}\(1 - \exp(-\alpha)\)
\end{equation}

\section{Numerical studies} \label{sec:numerical_studies}

To verify Eq.~(\ref{eq:measuredvar2}) numerically, we use a simple
Brownian dynamics \cite{ermak1978bdh} simulation of a bead fluctuating
in a harmonic potential. For each time step $\Delta t$, the change in
the bead position $\Delta x$ is given by a discretization of the
overdamped Langevin equation:
\begin{equation}\label{eq:bd_step}
\Delta x =  \frac{D}{k_B T} f_{\text{det}} + \delta x (\Delta t)
\end{equation}
where $\delta x (\Delta t)$ is a Gaussian random variable with
$\llangle \delta x \rrangle = 0$ and $\bigl\langle \( \delta x \)^2
\bigr\rangle = 2 D \Delta t$, and the deterministic force
$f_{\text{det}} = -k x$ corresponds to a harmonic potential as in our
calculation. Motion blur is simulated by time-averaging the simulated
bead positions over a finite integration time $W$. To minimize errors
due to discretization, the simulation sampling time is much smaller
than both $W$ and $\Gamma/m$.
\begin{figure}[tbp]
  \begin{center}
    \includegraphics[width=0.9\textwidth]
    {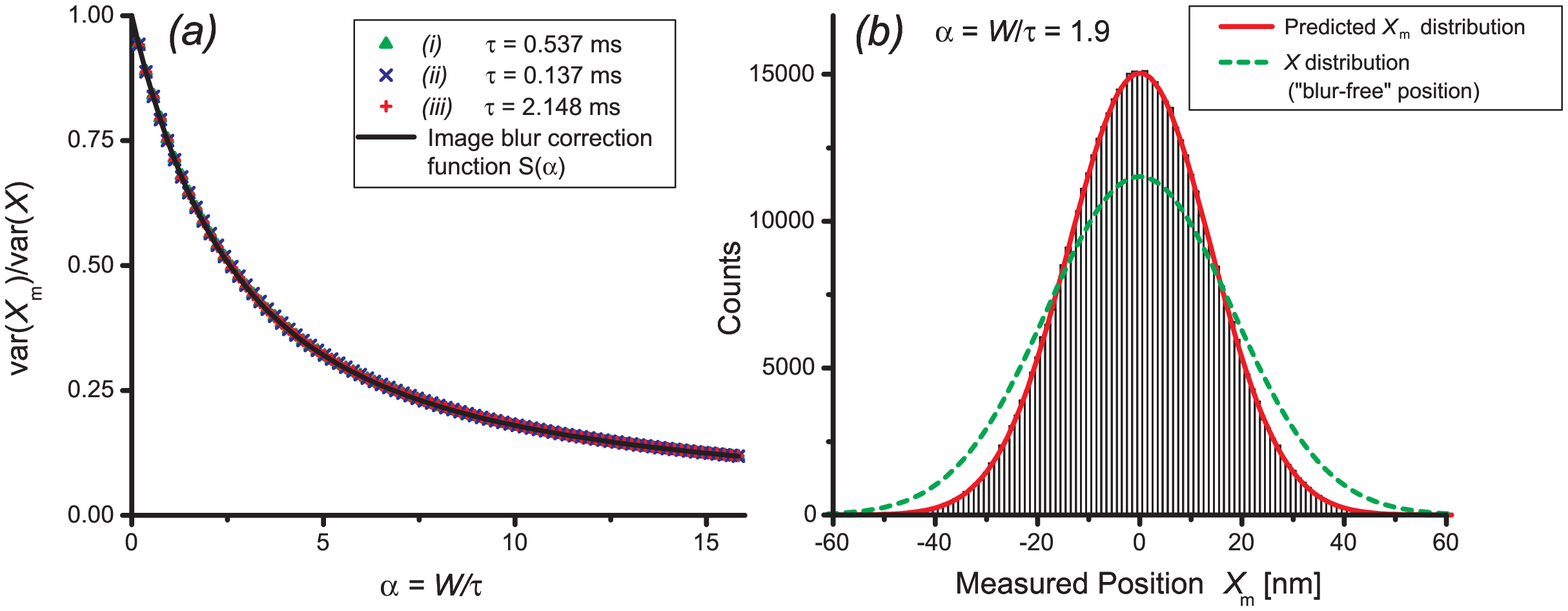}
    \caption{\label{fig:bdsim_universal_curve}(a) Brownian dynamics
    simulation results for measured variance as a function of exposure
    time. Data has been rescaled and plotted alongside $S(\alpha)$,
    the motion blur correction function of
    Eq.~(\ref{eq:universal_blur_fn}), showing excellent agreement
    within the expected error. The step size of the simulation is by
    $1 \mu \text{s}$, which is less than $0.01 \tau$ for all three
    simulations. The different simulation settings are: (i) $1.6 \:
    \mu\text{m}$ bead radius, $k = 0.05$ pN/nm, $\tau = 0.537 \:
    \text{ms}$, (ii) $0.4 \: \mu\text{m}$ bead radius, $k = 0.05$
    pN/nm, $\tau = 0.134 \: \text{ms}$, (iii) ($1.6 \: \mu\text{m}$
    bead radius, $k = 0.0125$ pN/nm, $\tau = 2.148 \: \text{ms}$) (b)
    Histogram of measured positions for simulation run (c) for an
    exposure time of 4 ms. It is a Gaussian distribution as expected
    \cite{wang1945tbm}. The normal curve with the predicted variance
    is superimposed showing excellent agreement. The expected
    distribution for an ideal ``blur-free'' measurement system is
    superimposed as a dotted line.}
  \end{center}
\end{figure}
Fig.~\ref{fig:bdsim_universal_curve}(a) shows the simulation results
for 3 different bead and spring constant settings as described in the
caption. Agreement with the motion blur correction function
$S(\alpha)$ of Eq.~(\ref{eq:universal_blur_fn}) is within the
fractional standard error of the variance, $\sim \sqrt{2/N}$
\cite{kenney1951msp}. We observe in Fig.
\ref{fig:bdsim_universal_curve}(b) that the distribution of measured
positions is a Gaussian random variable with variance $\text{var}(X_m)
< k_B T/k$, and is therefore fully characterized by the mean (trap
center) and the measured variance calculated in
Eq.~(\ref{eq:measuredvar2}).


\section{Experimental verification} \label{sec:experimental_verification}

\subsection{Instrument description}
The optical trap is formed by focusing 1064 nm near-IR laser light
(Coherent Compass 1064-4000M Nd:YVO$_4$ laser) through a high
numerical aperture oil immersion objective (Zeiss Plan Neofluar
100x/1.3) into a closed, water filled chamber. Laser power is varied
with a liquid-crystal power controller (Brockton Electro-Optics). This
optical tweezers system is integrated into an inverted light
microscope (Zeiss Axiovert S100).

The trapped bead is imaged with transmitted bright field illumination
provided by a 100 W halogen lamp (Zeiss HAL 100). The image is
observed with a high-speed cooled CCD camera with adjustable exposure
time (Cooke high performance SensiCam) connected to a computer running
custom data acquisition software \cite{heinrich_software1}. Each video
frame is processed in real-time to determine the position of the
trapped bead. Fast one-dimensional position detection is accomplished
by analyzing the intensity profile of a single line passing through
the bead center. To increase the signal to noise ratio and the frame
rate, the camera bins (i.e. spatially integrates) several lines about
the bead center (32 in this experiment) to form the single line used
in analysis. A third order polynomial is fit to the two minima
corresponding to the one-dimensional ``edges'' of the bead, giving
sub-pixel position detection with a measured accuracy of about 2
nm.

Tracking errors can be included in the measured variance by adding the
parameter $\varepsilon^2$ to Eq.~(\ref{eq:measuredvar2}), i.e.
\begin{equation}\label{eq:measuredvar_werror}
\text{var}(X_m) = \frac{2 k_B T}{k} \( \frac{\tau}{W} -
\frac{\tau^2}{W^2} (1 - \exp(-W/\tau)) \) + \varepsilon^2
\end{equation}
$\varepsilon^2$ can be interpreted as the measured variance of a
stationary particle, provided there are no correlations between the
tracking error and the measured positions. A detailed treatment of
such particle tracking errors is provided in reference
\cite{savin2005sad}, where the authors also stress the importance of
using identical conditions of noise and signal quality when comparing
the $\varepsilon^2$ parameter between different experimental
runs. Care was taken to achieve these conditions as is described
below.


\subsection{Experimental conditions} \label{subsec:experimental}
The sample chamber was prepared with pure water and polystyrene beads
(Duke Scientific certified size standards 4203A, 3.063 $\mu $m $\pm $
0.027 $\mu $m). Experiments were performed by holding a bead in the
optical trap and varying the power and the exposure time. The bead was
held 30 $\mu $m from the closest surface, and the lamp intensity was
varied with exposure time to ensure a similar intensity profile for
each test. This ensured that the noise and signal quality between
experimental runs was very similar, as validated by the results. For
each test, both edges of the bead in one dimension were recorded and
averaged to estimate the center position.

\subsection{Experimental Results} \label{sec:experimentalresults}

The one-dimensional variance of a single bead in an optical trap was
measured at various laser powers and exposure times. Low frequency
instrument drift was filtered out as described in Appendix
\ref{appendix:variance}. For each power, measured variance vs.\
exposure time data was fit with Eq.~(\ref{eq:measuredvar_werror}) to
yield values for the spring constant $k$, friction factor $\gamma$,
and tracking error $\varepsilon ^2$. Error estimates in the variance
were calculated from the standard error due to the finite sample size,
and variations due to vertical drift.

Error in the fitting parameters indicate that the best estimates for
$\gamma $ and $\varepsilon^2$ occur at the lowest and highest powers,
respectively. These estimates both agree within 2{\%} of the error
weighted average for all powers. For the nominal bead size, $\gamma$
agrees with the Stokes' formula calculation to within 11{\%},
indicating a slightly smaller bead or lower water viscosity than
expected.  Additionally, the estimate of tracking error $\varepsilon $
determined from the fit compares favorably with the standard deviation
in position of a stuck bead, differing by about half a nanometer.

For a single bead observed under identical measurement conditions,
$\gamma$ and $\varepsilon^2$ are expected to remain essentially
constant as the laser power is varied. Good consistency was found
between the determined values of $\varepsilon^2$ from different
experimental runs, due to the protocol of matching the signal strength
between tests. While laser heating could cause $\gamma$ to decrease
with increasing power, this effect should be small for the $<$ 500 mW
powers used here \cite{peterman2003lih, celliers2000mlh}, so this
effect was neglected.

\begin{figure}[htbp]
  \begin{center}
    \includegraphics[width=0.5\textwidth]
    {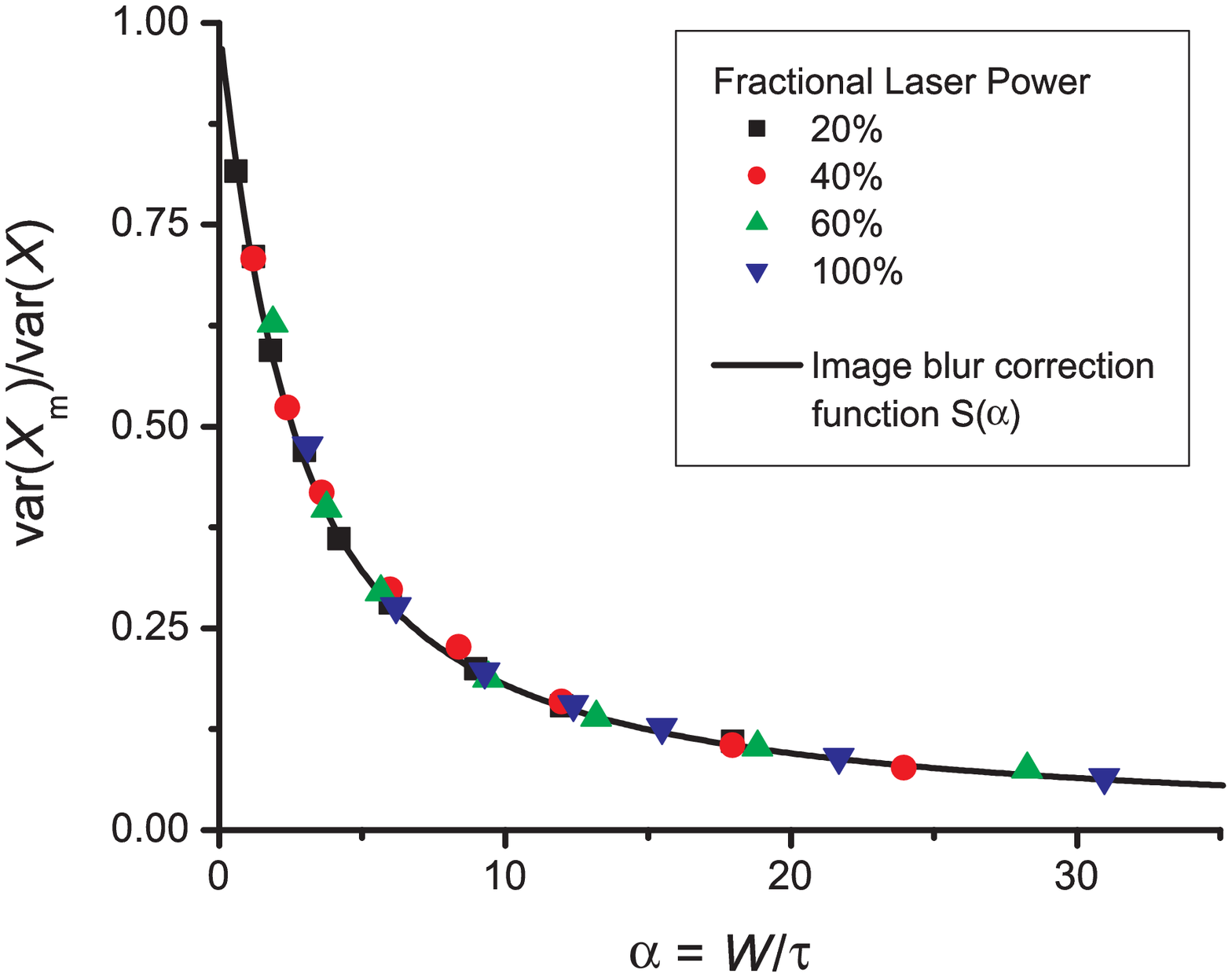}
    \caption{\label{fig:universal_curve_expt}Fractional variance
    $(\text{var}(X_m)/\text{var}(X))$ vs.\ dimensionless exposure time
    $\alpha = W/\tau$ for experimental optical trap data at 4
    different powers. Overlaid on the data is the motion blur
    correction function $S(\alpha)$ given by
    Eq.~(\ref{eq:universal_blur_fn}).}
  \end{center}
\end{figure}

\begin{figure}[htbp]
  \begin{center}
    \includegraphics[width=0.5\textwidth]
    {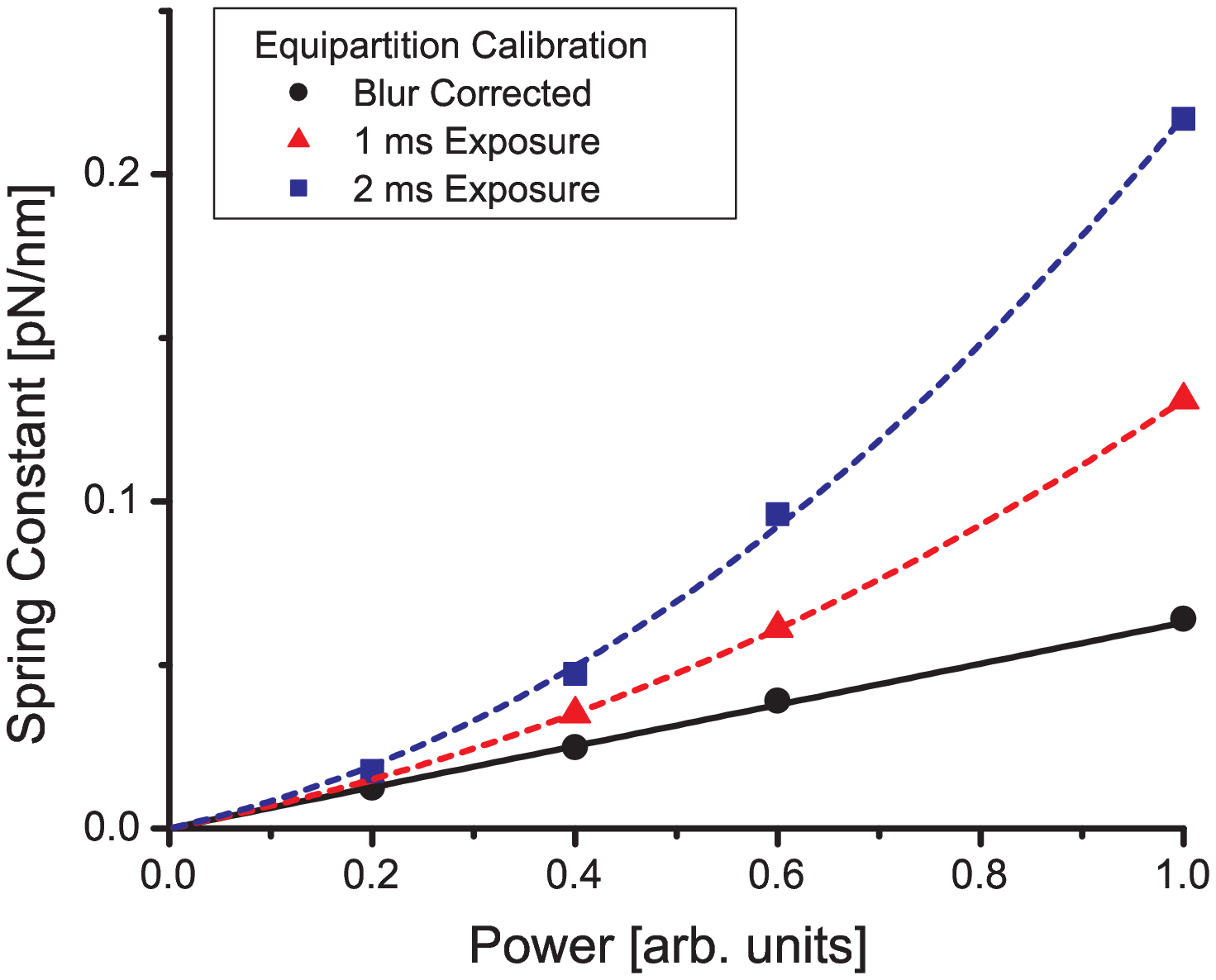}
    \caption{\label{fig:spring_constant_expt}Spring constant vs.\
    power for a single bead in the optical trap. The na\"\i ve
    equipartition measured spring constant with 1 ms and 2 ms exposure
    times (red triangles and blue squares, respectively) is compared
    with the blur corrected spring constant (black circles). The
    dashed blue and red lines going through the uncorrected data
    represent non-linear fits to the blur model assuming a linear
    relationship between $k$ and laser power, i.e. $k = c P$, as
    discussed in subsection\,(\ref{sec:diss_vary_k}). The values
    obtained from these fits for $c$ and $\gamma$ agree within error
    with the ``black circle'' values obtained by varying the exposure
    time.}
  \end{center}
\end{figure}

Holding $\gamma $ and $\varepsilon^2$ constant for all powers, the raw
data was re-fit with Eq.~(\ref{eq:measuredvar_werror}) to yield
$k$. The data for all 4 powers was error-corrected by subtracting
$\varepsilon^2$ and was rescaled according to
Eq.~(\ref{eq:measuredvar2}) and Eq.~(\ref{eq:universal_blur_fn}). This
non-dimensionalized data is plotted alongside the motion blur
correction function in Fig.~\ref{fig:universal_curve_expt}, showing
near-perfect quantitative agreement. This exceptional agreement
further validates our treatment of $\gamma $ and $\varepsilon^2$. A
plot of spring constant vs.\ dimensionless power is shown in
Fig.~\ref{fig:spring_constant_expt}, demonstrating the discrepancy
between the blur-corrected spring constant and na\"\i ve spring
constant for different integration times. Even for a modest spring
constant of 0.03 pN/nm and a reasonably fast exposure time of 1 ms,
the expected error is roughly 50{\%}. We also note that the
blur-corrected spring constant increases linearly with laser power as
expected from optical-trapping theory. Once confirmed for a given
system, this linearity can be exploited to determine not only the
spring constant as a function of power but also the diffusion
coefficient of the bead. This is discussed in
subsection\,(\ref{sec:diss_vary_k}), and presented in
Fig.~\ref{fig:spring_constant_expt}.

When the data acquisition rate is sufficiently high relative to $1/\tau
= k/\gamma$, it is feasible to calibrate the trap using the bead
position power spectrum, allowing comparisons to the previous results
at low laser power. Power spectrum fitting with the blur-corrected and
aliased expression (Eq.~(\ref{eq:powerspecaliased2})) at the lowest
power yielded both a spring constant and friction factor that agree
with the blur-corrected equipartition values to within 1{\%}.  Fits of
the same data using the na\"\i ve expression (Eq.
(\ref{eq:powerspecnaive}), not corrected for exposure time or
aliasing) provided slightly worse results, overestimating the spring
constant by 3{\%} and the friction factor by 7{\%}. (See Appendix
\ref{appendix:powerspec} for procedural details.)

For an additional check that does not rely on fluctuations, a purely
mechanical test was performed and compared with the corrected power
spectrum fit. This test consisted of a bead drop experiment to
determine the bead radius and friction factor, and a trap recoil
experiment to determine the spring constant. The bead drop was
performed by releasing a bead and recording its average velocity over
a known distance. The trap recoil experiment was performed by
measuring the exponential decay of the same bead as it returned to the
trap center after deviation in one dimension.  This mechanical test
agrees with the corrected power spectrum fit to within 5{\%} for the
determination of both the spring constant and the friction factor.

\section{Discussion: Practical suggestions for calibrating an optical trap}
\label{sec:discussion}

In this section, we present some practical techniques for measuring
the spring constant $k$ and diffusion coefficient $D$ of a
harmonically confined Brownian particle. We will assume that the
temperature $T$ is known. The approaches here are generic, and can be
used even if the confining potential is not an optical trap
(e.g. beads embedded in a gel, etc.) We will continue to treat the
measured position as an unweighted time average of the true position
over the integration time $W$ (Eq.~(\ref{eq:measured_bead_position})),
which is consistent with the experimental results for our detection
system. In other situations, e.g. if the rise and fall time are not
negligible relative to the exposure time, these equations and ideas
can be readily generalized as noted in
section\,(\ref{sec:calculation_section}).

\subsection{Determining $k$ from $D$ and $W$.}

If the diffusion coefficient $D$ of the confined particle and the
integration time $W$ of the instrument is known, the correct spring
constant $k$ can be directly obtained from the measured variance
$\text{var}(X_m)$ by using equation \ref{eq:measuredvar2}. If the
tracking error $\varepsilon$ is significant, $\varepsilon^2$ should
first be subtracted from the measured variance as in equation
\ref{eq:measuredvar_werror}. While we cannot in general isolate $k$ in
this transcendental equation, it can easily be found numerically by
utilizing a standard root-finding method. Alternatively, an
approximate closed form solution for $k$ is derived in Appendix
\ref{appendix:k_approx}.

\subsection{Determining $k$ and $D$ by varying $W$}

Even if the data acquisition rate of the system is not fast enough to
permit a blur-corrected power spectrum fit (as described in Appendix
\ref{appendix:powerspec}), $k$ and $D$ can still be determined by
measuring the variance at different shutter speeds and fitting to the
blur-corrected variance function. This technique is demonstrated in
the experimental results section, and yields accurate measurements
provided the integration time is not too much larger than the trap
relaxation time ($\alpha$ is not much larger than 1). Practically
speaking, this is a useful technique, as the maximum shutter speed of
a camera is often much faster than the maximum data acquisition speed
(e.g. it is much easier to obtain a video camera with a 0.1 ms shutter
speed than a camera with a frame rate of 10 kHz). Furthermore, this
approach for quantifying the power spectrum from the blur is quite
general, and could be used in other systems. As long as the form of
the power spectrum is known, the model parameters could be determined
by measuring the total variance over a suitable spectrum of shutter
speeds.

\subsection{Determining $k$ and $D$ by varying $k$}
\label{sec:diss_vary_k}

Other approaches are possible if the confinement of the particles can
be varied in a controlled way, i.e. by varying the laser power of the
optical trap. If the spring constant varies linearly with laser power,
(which is typically true and was confirmed for our system in
subsection\,(\ref{sec:experimentalresults})), the first observation is
that the spring constant only needs to be measured at a single laser
power, as it can be extrapolated to other laser powers. Typically
calibration should be done at a low power, as this usually increases
the accuracy of both the power spectrum fit and the blur correction
technique.

Linearity between the spring constant and laser power can be further
exploited to determine both $k$ and $D$ by measuring the variance of a
trapped bead at different laser powers but with the same shutter
speed. Such data can be fit to the blur model (equation
\ref{eq:measuredvar2}, recalling that $\alpha = W D k/(k_B T)$) by
introducing an additional fitting parameter $c$ that relates the laser
power $P$ to the spring constant, i.e. we make the substitution $k = c
P$, and perform a non-linear fit to $\text{var}(X_m)$ vs.\ power data
in order to determine $c$ and $D$.  Equivalently, we can express the
na\" \i ve spring constant $k_m = k_B T/\text{var}(X_m)$ as a function
of $c$ and $P$, and perform a fit to $k_m$ vs. $P$ data as shown in
Fig.~\ref{fig:spring_constant_expt} of the experimental results
subsection\,(\ref{sec:experimentalresults}), where the viability of
this method is demonstrated.

\subsection{Design strategies for using the blur technique}

When using these motion blur techniques to characterize the dynamics
of confined particles, we reiterate that it is the shutter speed and
not the data acquisition speed that limits the dynamic range of a
measurement. Thus, even inexpensive cameras with fast shutter speeds
can make dynamical measurements without requiring the investment of a
fast video camera. Alternative methods for controlling the exposure
time are the use of optical shutters or strobe lights.

\section{Conclusions}

We have experimentally verified a relationship between the measured
variance of a harmonically confined particle and the integration time
of the detection device. This yields a practical prescription for
calibrating an optical trap that corrects and extends both the
standard equipartition and power spectrum methods. By measuring the
variance at different shutter speeds or different laser powers, the
true spring constant can be determined by application of the motion
blur correction function of
Eq.~(\ref{eq:universal_blur_fn}). Additionally, this provides a new
technique for determining the diffusion coefficient of a confined
particle from time-averaged fluctuations.

The dramatic results from our experiment indicate that integration
time of the detection device cannot be overlooked, especially with
video detection.  Furthermore, we have shown that motion blur need not
be a detriment if it is well understood, as it provides useful
information about the dynamics of the system being studied.

\appendix

\section{Calculation of the measured variance of a harmonically trapped Brownian particle}\label{appendix:calculation_section}

In this appendix, a derivation of the motion blur correction function
Eq.~(\ref{eq:universal_blur_fn}) is presented. This quantifies how the
measured variance depends upon the spring constant $k$, the diffusion
coefficient of the particle $D$, and the integration time of the
imaging device $W$ (notation is as introduced in
section\,(\ref{sec:calculation_section})). The derivation follows
standard techniques (e.g. \cite{oppenheim1996ss}) and is similar to
calculations presented in references
\cite{yasuda1996dmt,savin2005sad,savin2005rfe,wang1945tbm}. For
completeness, we present the calculation in two different ways:
(\ref{sec:fourierspace}) a frequency-space calculation that convolves
the true particle trajectory with the appropriate moving-average
filter, and (\ref{sec:realspace}) a real-space calculation using
Green's functions. An expression for the modified power spectrum of
the harmonically confined bead that accounts for the effects of
filtering and aliasing is included in the frequency-space calculation
(see subsection\,(\ref{sub:blurcorpowerspec}))

\subsection{Frequency-space calculation} \label{sec:fourierspace}

The measured trajectory of a particle in the presence of motion blur
$X_m(t)$ can be calculated by convolving the true trajectory $X(t)$
with a rectangular function,
\begin{equation} \label{eq:conv}
X_{m}(t)= X(t) \ast H(t) \equiv \int X(t')H(t-t') \d[t']
\end{equation}
where $H(t)$ is defined by:
\begin{equation} \label{eq:filter}
H(t)=\left\{ \begin{array}{cl}
\frac{1}{W} & 0 < t \le W \\
0 & \text{elsewhere}
\end{array}\right.
\end{equation}
The integral is taken over the full range of values (i.e. $t'$ is
integrated from $-\infty$ to $+\infty$), which is our convention
whenever limits are not explicitly written. The width of the rectangle
$W$ is simply the integration time as previously defined.  This
convolution acts as an ideal moving average filter in time, and is
consistent with the integral expression for $X_m(t)$ given in
Eq.~(\ref{eq:measured_bead_position}).

Taking the power spectrum of Eq.~(\ref{eq:conv}) yields:
\begin{equation}\label{eq:powerspec}
P_m (\omega ) \equiv \left\vert \tilde{X}_m(\omega) \right\vert^2
=\left\vert \tilde{X}(\omega ) \right\vert^2\left\vert
\tilde{H}(\omega ) \right\vert^2
\end{equation}
Where the Fourier transform is denoted by a tilde,
e.g. $\tilde{X}(\omega) = \int X(t) \exp(i \omega t)\d[t]$, and
$\omega$ is the frequency in radians/second. The theoretical power
spectrum $P(\omega)$ is given by:
\begin{equation}\label{eq:powerspecnaive}
P(\omega )\equiv \left\vert \tilde{X}(\omega ) \right\vert^2
=\frac{2\gamma k_B T}{\gamma^2\omega^2 + k^2}
\end{equation}
where $\gamma$ is the friction factor of the particle, and is related
to the diffusion coefficient by the Einstein relation $\gamma = k_B
T/D$.  This power spectrum has been well-described previously
\cite{svoboda1994bao,wang1945tbm,gittes1998san}, and is derived in
section\,(\ref{sec:realspace}).

The power spectrum of the moving average filter can be expressed as a
squared sinc function:
\begin{equation} \label{eq:powerspech}
\left\vert \tilde{H}(\omega) \right\vert^2= \left( {\frac{\sin(\omega W/2)}{\omega W/2}} \right)^2
\end{equation}

Using Parseval's Theorem and integrating the power spectrum
$P(\omega)$ yields the true variance of $X(t)$,
\begin{equation} \label{eq:var_parseval}
\text{var}(X) =\frac{1}{2\pi}\int P(\omega) \d[\omega] =\frac{k_B
T}{k}
\end{equation}
which is in agreement with the equipartition theorem. Similarly, we
calculate the measured variance $\text{var}(X_m)$ as a function of the
exposure time $W$ and the friction factor $\gamma$ by integrating the
power spectrum of the measured position (Eq.~(\ref{eq:powerspec})):
\begin{eqnarray}\label{eq:measuredvarfourier}
\text{var}(X_m) &= &\frac{1}{2\pi}\int P_m(\omega) \d[\omega] \\
&=& \frac{2 k_B T}{k} \( \frac{\tau}{W} - \frac{\tau^2}{W^2} (1 -
\exp(-W/\tau)) \) \label{eq:measuredvarfourier2}
\end{eqnarray}
where $\tau = \gamma / k$, the trap relaxation time. Writing this
formula in terms of the dimensionless exposure time,
\begin{equation}\label{eq:dim_exp_def}
\alpha \equiv \frac{W}{\tau}
\end{equation}
and the variance of the true bead position $\text{var}(X) = k_B T/k$
  yields:
\begin{equation}\label{eq:measuredvar2_appendix}
\text{var}(X_m) = \text{var}(X) S(\alpha)
\end{equation}
where $S(\alpha)$ is the motion blur correction function:
\begin{equation}\label{eq:universal_blur_fn_appendix}
S(\alpha) = \frac{2}{\alpha} - \frac{2}{\alpha^2}\(1 -
\exp(-\alpha)\)
\end{equation}

\subsection{Blur-corrected filtered power spectrum} \label{sub:blurcorpowerspec}
Often, trap calibration is performed by fitting the power spectrum of
a confined particle. Here we provide a modification to the standard
functional form $P(\omega)$ that accounts for both exposure time
effects and aliasing. Combining expressions \ref{eq:powerspec},
\ref{eq:powerspecnaive} and \ref{eq:powerspech}, we can see the effect
of exposure time on the measured power spectrum:
\begin{equation} \label{eq:powerspecxm}
P_m (\omega )=\frac{2\gamma k_B T}{\gamma ^2\omega ^2+k^2}
\left( {\frac{\sin(\omega W/2)}{\omega W/2}} \right)^2
\end{equation}
Additionally, the effect of aliasing can be accounted for:
\begin{eqnarray} \label{eq:powerspecaliased}
P_{\text{aliased}} & =& \sum_{n = -\infty}^{+\infty} {P_m (\omega
  +n\omega _s )} \\
& = &\sum_{n = -\infty}^{+\infty} \frac{2\gamma k_B T}{\gamma ^2
  (\omega +n\omega _s)^2 + k^2}
\left( \frac{\sin((\omega + n \omega_s) W/2)}{(\omega + n \omega_s)
    W/2} \right)^2 \label{eq:powerspecaliased2}
\end{eqnarray}
where $\omega_s$ is the angular sampling frequency (i.e. the data
acquisition rate times $2 \pi$). Aliasing changes the shape of the
power spectrum, so neglecting it when fitting can cause errors. The
sum in Eq.~(\ref{eq:powerspecaliased2}) can be calculated numerically
and fit to experimental data. It is typically sufficient to calculate
only the first few terms.

It is important to note that aliasing does not affect our result for
the measured variance, Eq.~(\ref{eq:measuredvarfourier2}).  Aliasing
shifts power into the wrong frequencies, but does not change the
integral of the power. Hence, $\text{var}(X_m)$ is unchanged. A
detailed discussion of power spectrum calibration with an emphasis on
photodiode detection systems is given in reference
\cite{bergsoerensen2004psa}.

\subsection{Real-space calculation} \label{sec:realspace}

Since a Brownian particle follows a random trajectory $X(t)$, the
measured position $X_m$ is a random function of the true position of
the particle at the start of the integration time, i.e.
\begin{equation}\label{eq:measured_bead_position2}
X_m(x_0) = \frac{1}{W}\int_{0}^{W}X(t \mid x_0)\d[t]
\end{equation}
where $X(t \mid x_0)$ is the actual position of the bead at time $t$
given that it is at position $x_0$ at time zero, and $W$ is the
integration time as defined previously. In other words, even with
knowledge of the initial particle position, it is not possible to
predict what the measured position will be. However, the distribution
of $X_m$ is well-defined, and one can determine its moments.

The variance of the measured position is given by
\begin{equation}\label{eq:variancem}
\text{var}(X_m) \equiv \llangle X_m(X)^2 \rrangle - \llangle
X_m(X)\rrangle ^2
\end{equation}
Notice that to calculate the ensemble average $\llangle \dots
\rrangle$, we must average over both the random initial position $X$,
and the measured position for a given initial position $X_m(x)$.  For
the harmonic potential $U(x) = \frac{1}{2}k x^2$, $\llangle
X_m(X)\rrangle = 0$ by symmetry, so the variance reduces to
\begin{equation}\label{eq:variancem2}
\text{var}(X_m) = \int \rho_X(x_0) \ \llangle X_m(x_0)^2\rrangle \ \d[x_0]
\end{equation}
where $\rho_X(x_0)$ is the probability density of the initial
position, and the integral is taken over all space (consistent with
our previously stated convention). In equilibrium, $\rho_X(x_0)$ is
simply the Boltzmann distribution given in Eq.
(\ref{eq:boltzmann_weighting}).

Using Eq.~(\ref{eq:measured_bead_position2}), we express $\llangle
X_m(x_0)^2\rrangle$ as the double integral
\begin{align} \label{eq:avg_xm_x0}
\llangle X_m(x_0)^2\rrangle & =  \llangle \frac{1}{W^2}
\int_0^W\!\!\!\int_0^W X(t_1 \mid x_0) X(t_2 \mid x_0) \
\d[t_1]\d[t_2]\rrangle \\ & =  \frac{2}{W^2}
\int_0^W\!\!\!\int_0^{t_2} \llangle X(t_1 \mid x_0) X(t_2 \mid x_0)
\rrangle_{t_2 > t_1} \ \d[t_1]\d[t_2]
\end{align}
In the second step, the ensemble average is brought into the integral,
and the averaging condition $t_2 > t_1$ is added, which changes the
limits of integration.

The time-ordered auto-correlation function $\llangle X(t_1) X(t_2)
\rrangle_{t_2 > t_1}$ can be calculated using the Green's function of
the diffusion equation for a harmonic potential, $\rho\(x,t \mid
x_0,t_0\)$. The Green's function represents the probability density
for finding the particle at position $x$ at time $t$ given that it is
at $x_0$ at time $t_0$. It can be found by solving the diffusion
equation
\begin{equation}\label{eq:harmonic_diffusion}
\pdiff{\rho}{t} = D \pdiff[2]{\rho}{x} + \frac{D}{k_B T}
\pdiff{\rho}{x} k x + \frac{D}{k_B T} \rho k
\end{equation}
with the initial conditions $\rho(x,t_0) = \delta(x-x_0)$. The
solution to this problem is well-known \cite{doi1986tpd,wang1945tbm}
and is given by:
\begin{equation}\label{eq:harmonic_diffusion_gf}
\rho\(x,t \mid x_0,t_0\) = \frac{1}{\sqrt{2\pi k_B T V(t-t_0)/k}}
\exp\( - \frac{k \(x-x_0 \exp(-(t-t_0)/\tau)\)^2}{2 k_B T V(t-t_0)}\)
\end{equation}
where we have defined the dimensionless function:
\begin{equation}\label{eq:V}
V(t) = 1 - \exp(-2 t/ \tau)
\end{equation}
As before $\tau = \ k_BT/(kD) = \gamma / k$.  Notice that this is
simply a spreading Gaussian distribution with the mean given by the
deterministic (non-Brownian) position of a particle connected to a
spring in an overdamped environment, and with a variance that looks
like free diffusion at short time scales (i.e. initially increasing as
$2 D (t-t_0)$), but exponentially approaching the equilibrium value of
$k_B T/k$ on longer time scales.

The time-ordered auto-correlation function can be written as follows:
\begin{equation}\label{eq:autocorr}
\llangle X(t_1) X(t_2) \rrangle_{t_2 > t_1} = \int\!\!\!\int x_1 x_2
\rho(x_1,t_1 \mid x_0,0) \rho(x_2,t_2 \mid x_1,t_1) \d[x_1]\d[x_2]
\end{equation}
Putting in the Green's function of
 Eq.~(\ref{eq:harmonic_diffusion_gf}) and evaluating the integrals
 gives the result:
\begin{equation}\label{eq:autocorr2}
\llangle X(t_1) X(t_2) \rrangle_{t_2 > t_1} = x_0^2 \exp(-(t_2 +
t_1)/\tau) + \frac{k_B T \ V(t_1)}{k} \exp(-(t_2 - t_1)/\tau)
\end{equation}
where $V(t)$ and $\tau$ are as defined above.

Carrying out the double time integral in Eq.~(\ref{eq:avg_xm_x0}),
followed by the integral over the initial position $x_0$ of
Eq.~(\ref{eq:variancem2}) we obtain the final result for the measured
variance:
\begin{equation}\label{eq:measuredvar}
\text{var}(X_m) = \frac{2 k_B T}{k} \( \frac{\tau}{W} -
\frac{\tau^2}{W^2} (1 - \exp(-W/\tau)) \)
\end{equation}
This reproduces the result of the frequency-space calculation
presented in Eq.~(\ref{eq:measuredvarfourier2}).  The ideal power
spectrum can be obtained from the position auto-correlation function
of Eq.~(\ref{eq:autocorr2}). We determine the long-time limit of the
auto-correlation function by letting $t_1 \gg \tau$, which yields the
simplified equation:
\begin{equation}\label{eq:autocorr_longtime}
\llangle X(t_1) X(t_2) \rrangle = \frac{k_B T}{k} \exp(-\left\vert
t_2 - t_1\right\vert/\tau)
\end{equation}
Next, by taking the Fourier transform of this equation with respect to
$(t_2 - t_1)$ we obtain the standard result of
Eq.~(\ref{eq:powerspecnaive}).

\section{High-pass filtering in variance measurements}
\label{appendix:variance}
Calculation of the variance requires special attention, since low
frequency noise or drift can inflate the variance dramatically,
causing an underestimation of the spring constant. A high pass filter
can be used to remove low frequency noise, but the use of any ideal
filter lowers the variance by neglecting the contribution from the
removed frequencies (note Eq.~(\ref{eq:var_parseval})).

\begin{figure}[htbp]
  \begin{center}
    \includegraphics[width=0.5\textwidth]
    {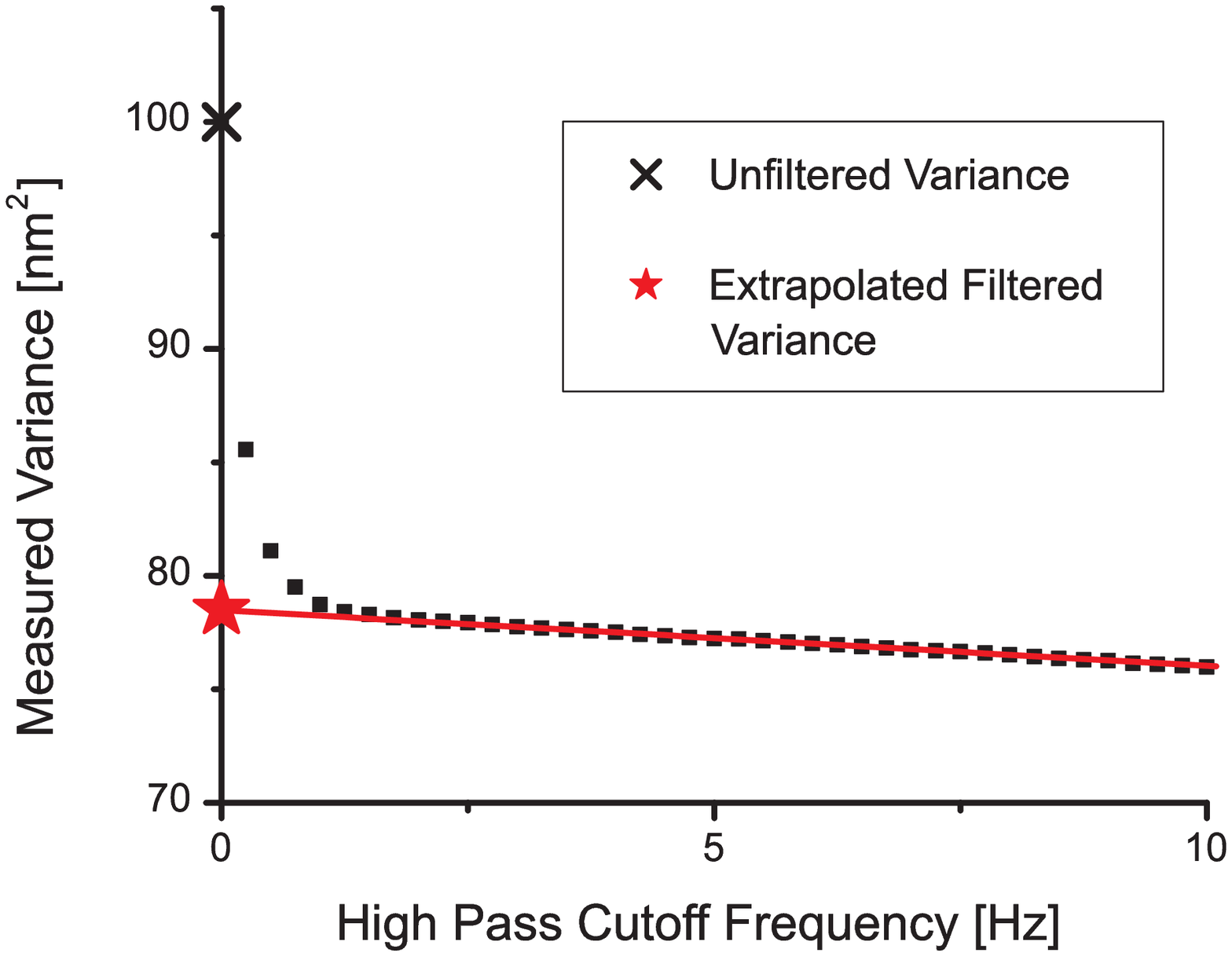}
    \caption{\label{fig:extrapolatedvar}Experimentally measured
    variance as a function of the high pass filter cutoff frequency
    shows a linear relation (line), which can be extrapolated to 0 Hz
    to reliably estimate the drift-free variance. The variance without
    filtering (cross) is $100 \: \text{nm}^2$, while the extrapolated
    variance (star) is $78.5 \: \text{nm}^2$}
  \end{center}
\end{figure}

To reliably estimate the variance while accounting for low frequency
drift, we first progressively high-pass filter the data over a range
of increasing cut-off frequencies. A plot of measured variance vs.\
cutoff frequency (Fig.~\ref{fig:extrapolatedvar}) clearly shows a
linear trend at frequencies below the corner frequency
(f$_{c}$=k/2$\pi \gamma )$. However, as the filtering frequencies
approach zero, drift causes the measured variance to increase beyond
its expected value. By applying a linear fit and extrapolating to the
0 Hz cutoff, we can reliably estimate the ``drift-free'' variance of
bead position.

\section{Experimental power spectrum calibration}
\label{appendix:powerspec}
\begin{figure}[htbp]
  \begin{center}
    \includegraphics[width=0.54\textwidth]
    {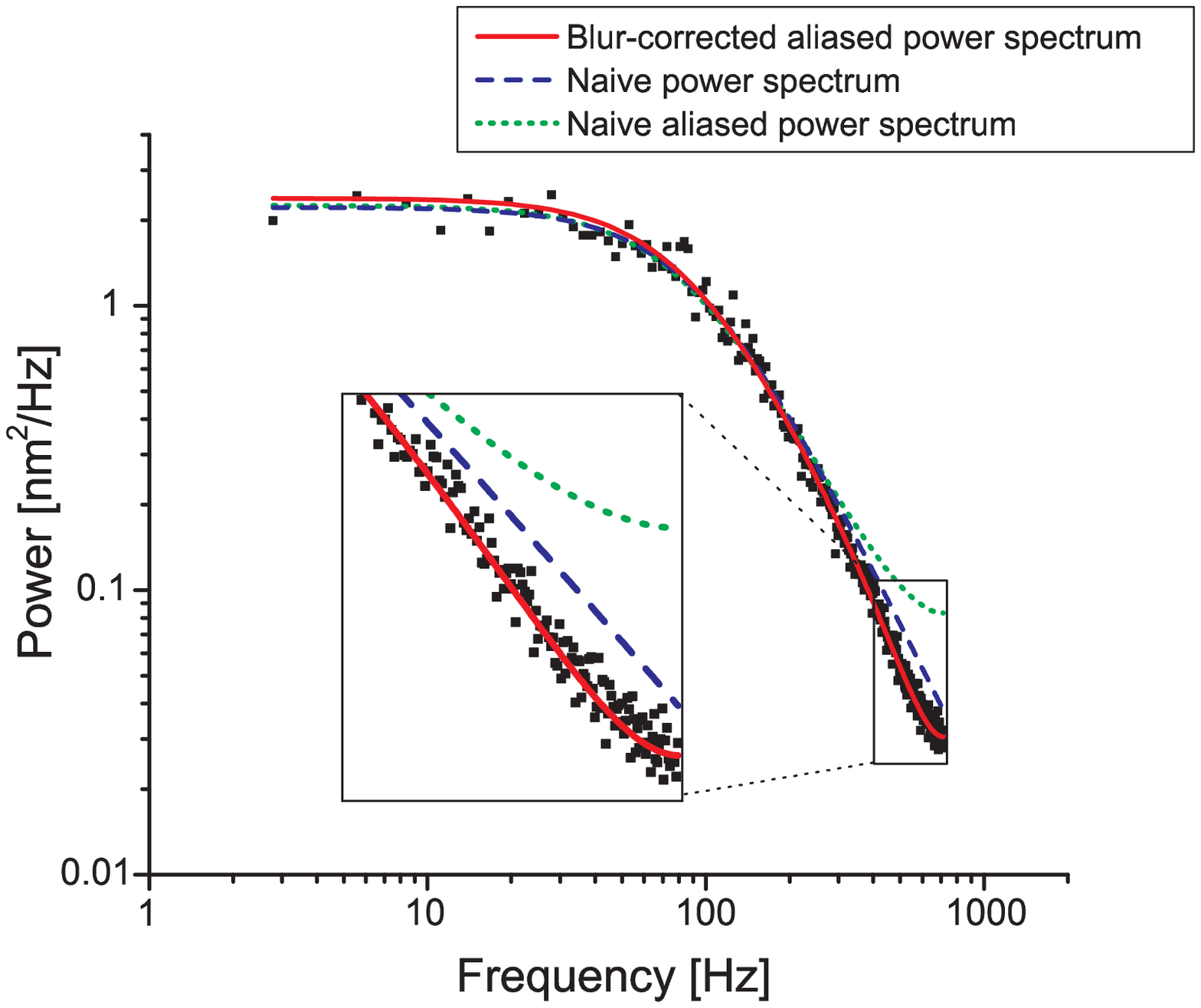}
    \caption{\label{fig:powerspec}A log-log plot of the one-sided
    power spectrum (dots) for a trapped bead, with theoretical models
    produced from a least squares fit to the data (blur-corrected and
    aliased, Eq.~(\ref{eq:powerspecaliased2}) solid line; na\"\i ve,
    Eq.~(\ref{eq:powerspecnaive}) blue dashed line; na\"\i ve aliased,
    green dotted line). The effect of the motion blur correction
    function $S(\alpha)$ is readily apparent from the clear
    discrepancy between the solid red and dotted green lines.}
  \end{center}
\end{figure}

\begin{figure}[htbp]
  \begin{center}
    \includegraphics[width=0.84\textwidth]
    {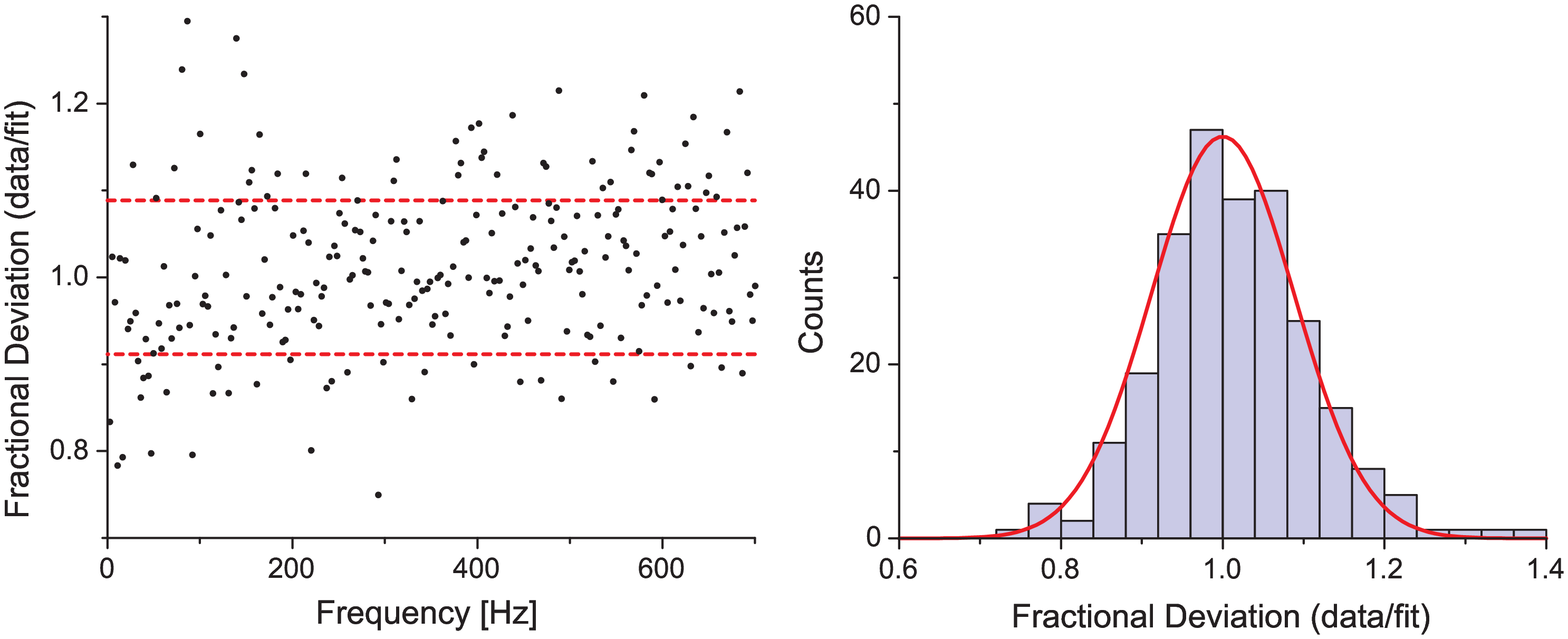}
    \caption{\label{fig:powerspec_residuals} Fractional deviation of
    the power spectrum data obtained by dividing the experimentally
    measured values (dots in Fig.~\ref{fig:powerspec}) by the fit
    obtained with the blur-corrected and aliased model (solid red line
    in Fig.~\ref{fig:powerspec}). \emph{Left:} Scatter plot
    demonstrating the quality of the fit; the two dashed red lines
    indicate the estimated standard deviation from unity of
    $1/\sqrt{128}$ \cite{bergsoerensen2004psa}. \emph{Right:}
    Histogram of the fractional deviation data overlaid with a
    Gaussian distribution with a standard deviation of $1/\sqrt{128}$
    (solid red line).}
  \end{center}
\end{figure}

Power spectrum calibrations were performed by fitting the one-sided
power spectrum with Eq.~(\ref{eq:powerspecaliased2}). The original
65536 data points taken at $\sim $1500 samples per second were blocked
into 128 non-overlapping segments. The power spectrum of the blocks
were calculated separately and averaged to produce the data in
Fig.~\ref{fig:powerspec}. This procedure is well described in the
literature \cite{gittes1998san,bergsoerensen2004psa}.  This data was
fit with the blur-corrected and aliased model of
Eq.~(\ref{eq:powerspecaliased2}) and compared with the commonly used
non-corrected power spectrum of Eq.~(\ref{eq:powerspecnaive}), both
with and without aliasing.  The quality of the fit to
Eq.~(\ref{eq:powerspecaliased2}) was further investigated by examining
the fractional deviation in the power (the measured data divided by
the model fit) as in reference \cite{bergsoerensen2004psa}. A scatter
plot and histogram of the fractional deviation is presented in
Fig.~\ref{fig:powerspec_residuals}. The histogram agrees well with a
Gaussian distribution with a standard deviation of $1/\sqrt{128}$ (see
reference \cite{bergsoerensen2004psa} for a thorough discussion of
power-spectrum fitting, including the expected scatter from unity).

Accounting for tracking error in the power spectrum fit is more
difficult than in the equipartition case, requiring knowledge of the
frequency dependence of the error. To investigate this in the current
study, the power spectrum of a stationary bead was subtracted from the
calibration power spectrum. We found that the fit parameters remained
practically unchanged (within 2{\%}), allowing us to neglect tracking
error in our power spectrum fits at low power.  It should be noted
that in other situations (e.g. different bead size or power),
modifications to the power spectrum due to tracking error could be
significant.

\section{Approximate analytical expression for $k$}
\label{appendix:k_approx}
When $W$ is not significantly larger than the trap relaxation time,
i.e. $\alpha = W/\tau = W k/\gamma$ is not much larger than 1, an
approximate version of equation \ref{eq:measuredvar2} can be
inverted to give a closed form solution for $k$.  First, we use a
Pad\'e approximation to express the motion blur correction function as:
\begin{equation}\label{padeapprox}
S(\alpha) \approx \frac{1 - 2\alpha/15 + \alpha^2/60}{1 + \alpha/5}
\end{equation}
Substituting this expression into equation
\ref{eq:measuredvar2} yields a quadratic equation that is
easily solved for $k$. This results in the following approximation for
the true spring constant:
\begin{equation}
k \approx \frac{30 \, k_B T}{2 D W +15 \, \text{var}(X_m) + \left[225
   \, \text{var}(X_m)^2 + 240 D W \text{var}(X_m) - 11 D^2
   W^2\right]^{1/2}}
\end{equation}
The Pad\'e approximation is good to within 3\% for $\alpha < 3$, which
corresponds to a blur correction factor of $S(3)\doteq 0.46$. In other
words, if the uncorrected equipartition method gives a spring constant
which is within a factor of 2 of the true value, this approximation
formula should be accurate to within 3\%, as we have tested
numerically.





\section*{Acknowledgments}

The authors would like to thank Evan Evans (Departments of Biomedical
Engineering and Physics, Boston University; Departments of Physics and
Astronomy and Pathology, University of British Columbia) for
scientific and financial support, providing the necessary laboratory
resources for this project, and for useful scientific advice and
discussions throughout.

The authors would like to thank Volkmar Heinrich (Department of
Biomedical Engineering, University of California, Davis) for initial
discussions which helped to catalyze this project, assistance with
building the optical trap including writing the data acquisition
software, and for useful scientific advice and discussions throughout.

In addition, the authors would like to thank the following people for
helpful discussions and feedback on the manuscript: Michael Forbes,
Ludwig Mathey, Ari Turner, and the anonymous reviewers of this
submission.

This work was supported by USPHS grant HL65333 from the National
Institutes of Health.


\begin{thebibliography}{10}
\bibitem{heinrich_software1}
Data aquisition software was written by Volkmar Heinrich.
\end{thebibliography}

\end{document}